\documentclass{iopart}

\usepackage{graphicx}
\usepackage{dcolumn}
\usepackage{bm}

\begin{document}

\title[The Chaotic Ball]
{The ``Chaotic Ball'' model, local realism and the Bell test loopholes}

\author{Caroline H Thompson$^1$\footnote{Email: ch.thompson1@virgin.net}  and Horst Holstein$^2$}
\address{$^1$ 11 Parc Ffos, Ffos-y-Ffin, Aberaeron, SA46 0HS, U.K.}
\address{$^2$ Department of Computer Science, 
	University of Wales, Aberystwyth, SY23 3DB, U.K.}

\date{\today}

\begin{abstract}
It has long been known that the ``detection'' or ``fair sampling'' loophole
could open the way for alternative ``local realist'' explanations for the 
violation of Bell tests.  It is usual, though, to assume fair sampling, so
that the loophole can be ignored.  We describe a model, along the same lines as
Pearle's of 1970 but considerably simpler, to illustrate intuitively why this 
may not be justified.  Two versions of the Bell test --- the standard one of 
form $-2 <= S <= 2$ and the currently-popular ``visibility'' test --- are at 
grave risk of bias.  Statements implying that experimental evidence ``refutes 
local realism'' or shows that the quantum world really is ``weird'' should be 
reviewed.  The detection loophole is on its own unlikely to account for more 
than one or two test violations, but when taken in conjunction with other 
loopholes (briefly discussed) it is seen that the experiments refute only a 
narrow class of ``local hidden variable'' models, applicable to idealised 
situations, not to the real world. The full class of local realist models may 
provide straightforward explanations not only for the publicised Bell-test 
violations but also for some lesser-known ``anomalies''.
\end{abstract}

\submitto{\JOB}

\noindent{\it Keywords\/ Local realism, hidden variables, Bell tests, bias, 
fair sampling, loopholes}

\maketitle 

\section{Introduction}
Pearle (1970~\cite{Pearle:1970}) showed that failure to detect some 
particles during Bell test experiments can allow local realist (hidden 
variable) explanations to reproduce almost exactly the quantum-mechanical 
(QM) predictions. This fact, which has become known as the ``detection'', 
``efficiency'' or ``fair sampling'' loophole, has been re-discovered many 
times~\cite{Fine:1982,Greenberger:1995,Gilbert:1996,Gisin:1999}, for 
Pearle's paper did not attract the wide publicity it deserved. It is cited 
by Clauser and Shimony in their review article (1978~\cite{Clauser:1978}) 
but by relatively few others. There seems, however, to have 
been general acceptance in the 1970's that ``fair sampling'' could not 
legitimately be assumed. Experimenters such as Freedman and Clauser in 
1972~\cite{Freedman:1972} and Fry and Thompson in 1976~\cite{Fry:1976} 
recognised that neither Bell's original test~\cite{Bell:1964} nor the 
variant introduced in 1969 by Clauser, Horne, Shimony and 
Holt~\cite{Clauser:1969} was appropriate to real experimental conditions. 
They used instead even later versions, described by Clauser and Horne in 
1974 (CH74)~\cite{Clauser:1974}, that were not dependent upon near-perfect 
detection efficiency. ``Normalisation'' in these modified tests was by 
comparison with coincidence counts obtained with the polarisers removed, 
not, as in the modern interpretation of Clauser {\it et al}'s 1969 test 
(hereafter referred to as the CHSH69 test\footnote{
The term ``CHSH69 test'' will be used in the current paper with its commonly 
accepted meaning, in which the denominator in the estimate for each ``correlation'' 
term is the total number of observed coincidences. Careful reading of the 
1969 paper reveals, however, that the authors had no intention of applying the 
inequality in this way. They describe how to use it in an experimentally 
realisable manner, which means effectively using the CH74 test. At no point do 
they mention the possibility of dividing by the total number of observed coincidences.
})
with the total when both were present. 

Clauser {\it et al}'s paper of 1969 had been inspired by Bell's original work. Clauser and 
Horne's of 1974 took account 
of his later ideas~\cite{Bell:1971}, making substantial 
improvements\footnote{
Bell kept consistently to a model and notation that 
were adapted to a spin-1/2 experiment, in which the ``outcome'' on side A 
of the experiment was denoted by {\it A}. In the 1964 paper this could 
take only values +1 and --1. By 1971 zero outcomes were also admitted. 
However, this notation --- used in all popular accounts of Bell's 
inequalities --- was both confusing, with the symbol {\it P} used for 
quantum correlation instead of for a probability, and unsuitable for use 
in optical applications. Clauser and Horne in 1974 broke away from this 
tradition. They extended Bell's 1971 idea to establish a new concept of 
hidden variable, one that did not determine the outcome but only its 
probability. They considered in the first instance a setup using plane 
polarised light and polarisers that had only one output, and derived 
in a very straightforward manner an inequality restricting the probability 
of coincidence instead of the quantum correlation. The latter, though 
this was not explicitly stated, is not in fact well defined for optical 
work, in that it does not allow for ``double detections'' --- the 
simultaneous occurrence of `+' and `--' values at the two outputs of 
a (two-channel) polariser.}
 and suggesting a test that did not 
demand high detector efficiency (instead, the assumption of ``no enhancement''
was made --- see section \ref{no enhancement} below).  In addition, detailed 
footnotes covered a wide range of other potential ``loopholes'' in real optical 
experiments. It is the 1969 paper, though, that is most frequently quoted and, 
since about 1980, the CHSH69 test and the almost-equivalent ``visibility'' test\footnote{
The ``visibility'' test used in several recent experiments depends 
on the fact that in certain conditions the CH74 test reduces to 
{\it v} $\le $ 1/$\surd $2 $\approx $ 0.71, where {\it v} is the visibility, 
$(max - min)/(max + min)$, of the coincidence curve. Its validity depends, 
among other things, on the assumption that this curve is truly sinusoidal.
} 
have come into favour. The assumption of ``fair sampling'' 
is accepted either with no comment at all or described by terms such as 
``plausible'', the hidden variable theories associated with its failure being 
dismissed as ``bizarre'' or requiring ``conspiracies'' between the 
detectors~\cite{Laloe:2001}. 

Some experimenters, for example N. and B. Gisin~\cite{Gisin:1999}, have 
taken due note of the importance of the detection loophole, but others seem 
unaware of its implications. The CHSH69 and visibility tests can only safely 
be used if the detection rates are very high --- a condition that has rarely 
if ever been met in practice. The trapped ion experiment of Rowe 
{\it et al.}~\cite{Rowe:2001}, acclaimed in some popular 
accounts~\cite{PhysWorld:2001} as having closed the final loophole, did
 indeed have very high efficiency, but it failed another crucial test: the 
ions could not legitimately be described as ``separated'' (they were in the 
same trap, controlled by the same laser), so Bell's inequality was not applicable.

The present paper, investigating conditions in which sampling is {\it not }
``fair'', is based on the ``Chaotic Ball'' model devised by C. H. Thompson 
and first published in 1996~\cite{Thompson:1996}. It has now been improved by the 
addition of an analytical formula derived by one of us (HH), enabling 
readers to explore its possibilities for themselves. The logic is 
essentially the same as Pearle's, but the geometry is very much 
simpler, aiming to illustrate principles rather than to reproduce the exact 
quantum-mechanical prediction.  (For an excellent illustration of Pearle's geometry see 
ref.~\cite{Risco-Delgado:1993}.) 

The model is most appropriate for the thought experiments of 
Bohm~\cite{Bohm:1951}, involving spin-1/2 particles, in which the direction 
of spin is taken to be a vector and used as the hidden variable. No claim is 
made, however, that it corresponds to the actual physics of Bohm's 
experiment, which has, it must be emphasised, never been performed and might 
not in practice ever be feasible. 

The majority of actual Bell test experiments have involved light, whose 
particle nature is debatable~\cite{Lamb:1995}. Some, for example Aspect's 
well-known experiments of 1981-2~\cite{Aspect:1981,Aspect:1982a,Aspect:1982b}, 
involved the polarisation of light; other more recent ones, for example 
Tapster~{\it et al}, 1994~\cite{Tapster:1994}, involved its phase and momentum. 
The basic local realist model that covers them all --- or, at least, those for 
which emitted pairs can be identified on detection without ambiguity --- predicts the 
probability of a coincidence to be
\begin{equation}
\label{eq1}
P(a,b)=\int\limits^{\Lambda} {d\lambda } \rho (\lambda )p_a (\lambda )p_b 
(\lambda ),
\end{equation}
where $\Lambda $ is the space spanned by the ``hidden variable'' $\lambda $, 
$a$ and $b$ are the detector settings and $p_{a}$ and $p_{b}$  functions of $\lambda 
$ giving the probabilities of detection at the two detectors. Some readers 
may prefer to think directly in terms of this formula, following Marshall, 
Selleri and Santos~\cite{Marshall:1983} or other articles by C. H. 
Thompson~\cite{Thompson:1997,Thompson:1999a}. The points raised 
in the current paper, though, are quite general, especially regarding the 
existence of less well known loopholes and the fact that not all Bell tests 
are identical or involve the same assumptions.

The Chaotic Ball, covering just one of the loopholes, corresponds in its basic
form to a rotationally invariant deterministic case of expression (\ref{eq1}), 
with $\rho$ constant and $p_{a}$ and $p_{b}$ taking only 
values 0 or 1. The geometrical difference between spin (for which 
``opposite'' means differing by $180^{\circ}$) and polarisation (for which it 
means differing by $90^{\circ}$) is of no significance so far as the principle 
illustrated is concerned. Even the apparently fundamental difference between 
a true particle (which can go only to one or other detector) and light 
(which may well, despite the claims of quantum opticians~\cite{Holbrow:2002,Grangier:1996}, 
go to both at once~\cite{Marshall:1996}) does not seriously affect the logic.

\section{The ``Chaotic Ball''}
Let us consider Bohm's thought experiment, commonly taken as the standard 
example of the entanglement conundrum that Einstein, Podolsky and Rosen 
discussed in their seminal 1935 paper~\cite{Einstein:1935}. A molecule is 
assumed to split into two atoms, $A$ and $B$, of opposite spin, that separate in 
opposite directions. They are sent to pairs of ``Stern-Gerlach'' magnets, 
whose orientations can be chosen by the experimenter, and counts taken of 
the various ``coincidences'' of spin ``up'' and spin ``down''. The obvious 
``realist'' assumption is that each atom leaves the source with its own 
well-defined spin (a vector pointing in any direction), and it is the fact 
that the spins are opposite that accounts for the observed coincidence 
pattern. (The realist notion of spin cannot be the same as the quantum 
theory one, since in quantum theory ``up'' and ``down'' are concepts defined 
with respect to the magnet orientations, which can be varied. Under quantum 
mechanics, the particles exist in a superposition of up and down states 
until measured.)

Bell's original inequality was designed to apply to the estimated ``quantum 
correlation\footnote{
The definition that Bell gave (page 15 of ref. \cite{Bell:1987}) for 
quantum correlation was the ``expectation'' value of the product of the 
``outcomes'' on the two sides, where the ``outcome'' is defined to be +1 
or --1 according to which of two possible cases is observed. It is to be 
assumed that he was using the word ``expectation'' in its usual statistical 
sense and that an unbiased estimate would be used.
}'' between the particles. He proved that the realist assumption, based on the premise that the detection events for a given pair of particles are independent, leads to statistical limits on this 
correlation that are exceeded by the QM prediction. He did not, however, 
specify how it was to be estimated in cases where not all the particles were 
detected.

\begin{figure}[htbp]
\centerline{\includegraphics[width=3.0in,height=1.67in]{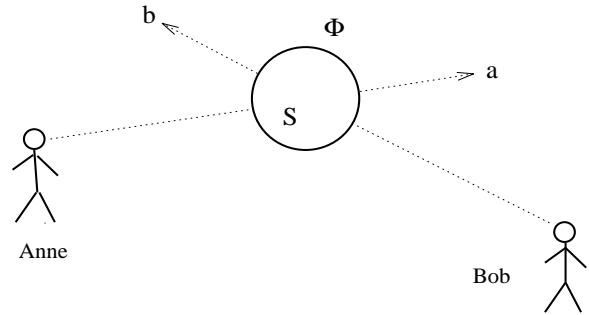}}
\caption{Anne, Bob and the Chaotic Ball.
 The letter $S$ is visible while $N$, opposite to it, is out of sight. 
$a$ and $b$ are directions in which the assistants are viewing the ball; 
$\Phi$ the angle between them.}
\label{fig1}
\end{figure}

When detection is perfect there is no problem, but when it is not, the 
``detection loophole'' creeps in. What assumptions can we reasonably make? 
Under quantum theory, the most natural one is that all emitted particles 
have an equal chance of non-detection (the sample detected is ``fair'', not 
varying with the settings of the detectors). The realist picture, however, 
is different.

Let us replace the detectors by two assistants, Anne (A) and Bob (B), the 
source of particles by a large ball on which are marked, at opposite points 
on the surface, an $N$ and an $S$ (Fig.~\ref{fig1}). The assistants look at 
the ball, which turns randomly about its centre (the term ``chaotic'', 
though bearing little relation to the modern use of the term, is retained 
for historical reasons). They record, at agreed times, whether they see 
an $N$ or an $S$. When sufficient records have been made they get together 
and compile a list of the coincidences --- the numbers of occurrences 
of $NN$, $SS$, $NS$ and $SN$, where the first letter is Anne's and the 
second Bob's observation.

The astute reader will notice that, if the vector from $S$ to $N$ corresponds to 
the ``spin'' of the atom, the model covers the case in which the spins on 
the $A$ and $B$ sides are {\it identical}, not opposite. Anne and Bob are 
looking at identical copies of the ball, which can conveniently be represented 
as a single one. This simplification aids visualisation whilst having no significant 
effect on the logic. The difference mathematically is just a matter of change of 
sign, with no effect on numerical values. In point of fact, the assumption of 
identical spins makes the model better suited to some of the actual optical 
experiments. Aspect's, for example, involved plane-polarised ``photons'' 
(not, incidentally, circularly polarised, as frequently reported) with parallel, 
not orthogonal, polarisation directions.

\begin{figure}[htbp]
\centerline{\includegraphics[width=2.0in,height=1.65in]{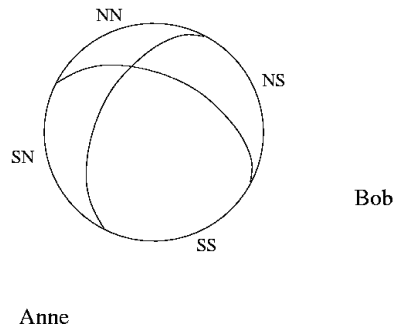}}
\caption{The registered coincidences: Chaotic Ball with perfect detectors.
The first letter of each pair denotes what Anne records, the second Bob, when 
the $S$ is in the region indicated.}
\label{fig2}
\end{figure}

With this simplification, geometry dictates that if the ball takes up all 
possible orientations with equal frequency (there is rotational 
invariance) then the relative frequencies of the four different 
coincidence types will correspond to four areas on the surface of an 
abstract fixed sphere as shown in Fig.~\ref{fig2}.

\begin{figure}[htbp]
\centerline{\includegraphics[width=3.0in,height=1.75in]{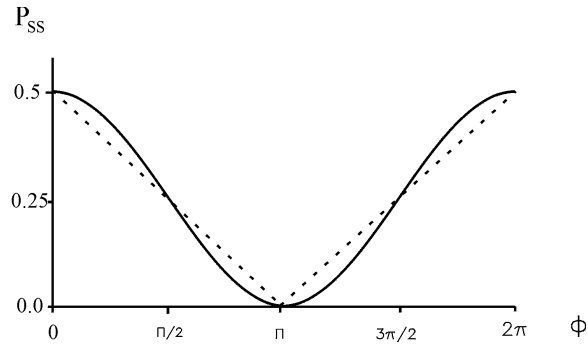}}
\caption{Predicted coincidence curves.
Dotted lines give the local realist prediction for the probability that, if there
are no missing bands, both Anne and Bob see an $S$; the curve is the QM prediction, 
$\frac{1}{2} \cos^2(\Phi /2)$.}
\label{fig3}
\end{figure}

Anne's observations correspond to two hemispheres, Bob's to a different 
pair, the dividing circles being determined by the positions of the 
assistants. We conduct a series of experiments, each with fixed lines of 
sight (``detector settings'') {\bf {\it a}} and {\bf {\it b}}. It 
can readily be verified that the model will reproduce the standard 
``deterministic local realist'' prediction, with linear relationship between 
the number of coincidences and $\Phi$, the angle between the 
settings\footnote{
The prediction of a linear relationship for the ``perfect'' 
case is most easily verified by drawing diagrams of the ball as seen from above. 
The dividing circles are then straight lines through the centre and the areas 
required are proportional to the angles between them.}. This is shown in Fig.~\ref{fig3}, which also shows the quantum mechanical prediction, a sine curve.

What happens, though, if the assistants do not both make a record at every 
agreed time? If the only reason they miss a record is that they are very 
easily distracted, this poses little problem. So long as the probability of 
non-detection can be taken to be random, the expected pattern of 
coincidences will remain unaltered. What, though, if the reason for the missing 
record varies with the orientation of the ball --- with the 
``hidden variable'', $\lambda$, the vector from $S$ to $N$?  
As mentioned in the literature~\cite{Feynman:1978}, in actual Stern-Gerlach experiments
some particles escape detection, failing to be deflected by either magnet.  
Could it be that these tend to be ones for which the vector had only very small 
components in the relvant ``up'' or ``down'' directions?

\begin{figure}[htbp]
\centerline{\includegraphics[width=3.0in,height=2.49in]{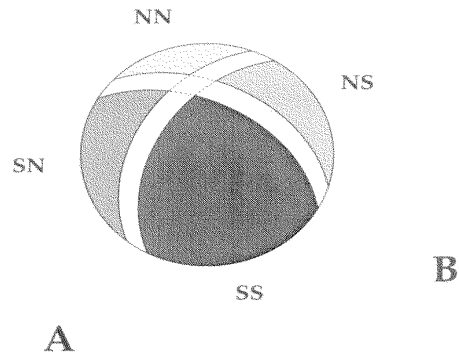}}
\caption{Chaotic Ball with missing bands.
There is no coincidence unless both assistants make a record, so some data is thrown away.}
\label{fig4}
\end{figure}

Suppose the ball is so large that the assistants cannot see the whole of the 
hemisphere nearest to them. The picture changes to that shown in Fig.~\ref{fig4}, in which 
the shaded areas represent the regions in which, when occupied by the $S$, coincidences 
will be recorded as indicated. The ratios between the areas, which are what matter in 
Bell tests, change --- indeed, some areas may disappear altogether. If the missing bands 
are {\it very} large, there will be certain positions of the assistants for which the 
estimated quantum correlation ($E$, equation (\ref{eq1}) below) is not even defined, 
since there are {\it no} coincidences. 

New decisions are required. Whereas before it was clear that if we wanted to 
normalise our coincidence rates we would divide by the total number of 
observations, which would correspond to the area of the whole surface, there 
is now a temptation to divide instead by the total shaded area. The former 
is correct if we want the proportion of coincidences to emitted pairs, but 
it is, regrettably, the latter that has been chosen in actual Bell test 
experiments. It is easily shown that the model will now inevitably, for a 
range of parameter choices, infringe the relevant Bell test if our estimates 
of ``quantum correlation'' are the usual ones, namely,
\begin{equation}
\label{eq2}
E(a,b)=\frac{NN+SS-NS-SN}{NN+SS+NS+SN},
\end{equation}
where the terms {\it NN} {\it etc}. stand for counts of coincidences in a self-evident 
manner.

\begin{figure}[htbp]
\centerline{\includegraphics[width=2.70in,height=2.28in]{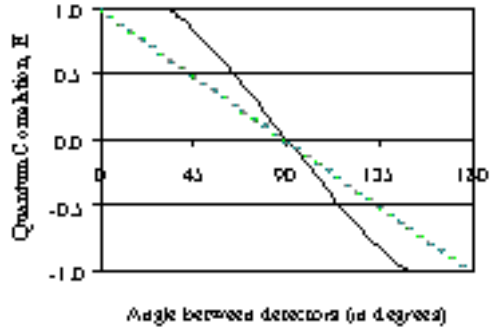}}
\caption{Predicted quantum correlation $E$ versus angle.
The curve corresponds to (moderate-sized) missing bands, the dotted line to 
none. See equation (4) of the text for the formula for the central section 
of the curve.}
\label{fig5}
\end{figure}

The Bell test in question is the CHSH69 test referred to above. It takes the 
form $-2 \le S \le 2$, where the test statistic is
\begin{equation}
\label{eq3}
S=E(a,b)-E(a,b')+E(a',b)+E(a',b').
\end{equation}
The parameters $a$, $a'$, $b$ and $b'$ are the detector settings: to evaluate the four terms 
four separate sub-experiments are needed. The settings chosen for the Bell 
test are those that produce the greatest difference between the QM and 
standard local realist predictions, namely $a$ = 0, $a'= \pi $/2, $b=\pi $/4 
and $b'$ = 3$\pi $/4. Since we are assuming rotational invariance, the value of 
$E$ does not depend on the individual values of the parameters but on their 
difference, $\Phi = b - a$, which is $\pi $/4 for three of the terms and 3$\pi 
$/4 for the fourth. We can therefore immediately read off the required 
values from a graph such as that of Fig.~\ref{fig5}, where the curve is calculated 
from the geometry of Fig. 4 (see next section).

When there are no missing bands it is clear that the numerical value of each 
term is 0.5 and that they are all positive. Thus with no missing bands the 
model shows that we have exact equality, with $S$ actually equalling 2.

If we do have missing bands, however, although the four terms are still all 
equal and all positive, each will have increased! {\it The Bell test will be infringed}.

An ``imperfection'' has {\it increased} the correlation, in contradiction to the opinion, 
voiced among others by Bell himself, that imperfections are unlikely ever to 
do this. It is not hard to imagine real situations in which something like 
these missing bands will occur (see earlier mention of the possibility of some particles
failing to be deflected in either direction), biasing this version of Bell's 
test in favour of quantum mechanics. Note that the ``visibility'' test used in more
recent experiments such as Tittel's long-distance Bell tests~\cite{Tittel:1997} 
is equally unsatisfactory, biased from the same cause.  As our model readily
shows, the ``realist'' upper limit on the standard test statistic when there 
is imperfect detection is 4, not 2, well above the quantum-mechanical one of 
$2\surd 2 \approx 2.8$.  The visibility can be as high as 1, not limited to the 
maximum of 0.5 that follows from the commonly-accepted assumptions.

\section{Detailed Predictions for the basic model}

\begin{figure}[htbp]
\centerline{\includegraphics[width=3.0in,height=3.0in]{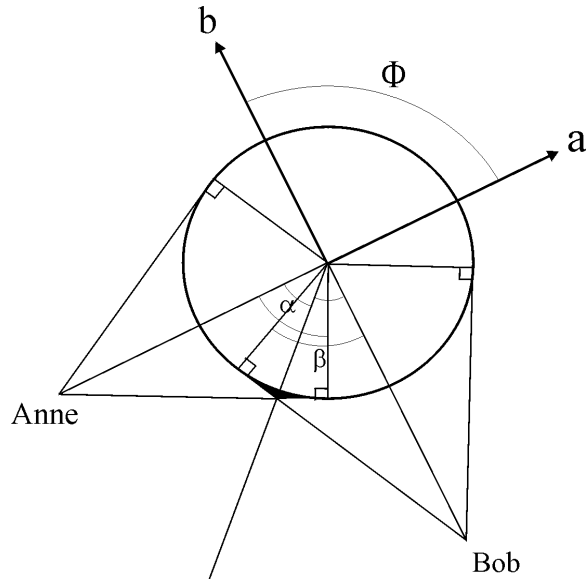}}
\caption{Definition of angles used in equation (4).}
\label{fig6}
\end{figure}

\begin{figure}[htbp]
\centerline{\includegraphics[width=2.8in,height=2.0in]{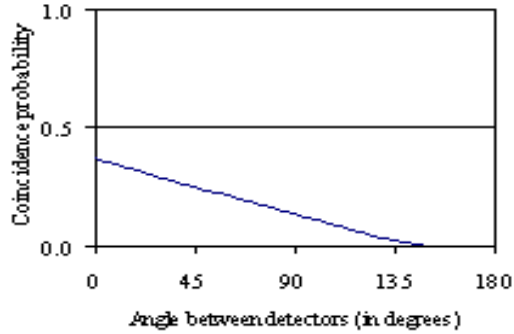}}
\caption{Predicted coincidence rate $P_{SS}$ for $\beta $ set at $75^{\circ}$ ($5\pi/12$).}
\label{fig7}
\end{figure}

The main formula for the proportion $P_{SS}$ of ``like'' coincidences such as 
{\it SS }with respect to the number of emitted pairs $N$ comes from the area of 
overlap of two equal-sized circles\footnote{We model here the simplest case, in 
which the two assistants stand at equal distances from the ball.} on the surface 
of a sphere (see Figs.~\ref{fig4} and~\ref{fig6}). It can be shown to be
\begin{equation}
\label{eq4}
P_{SS} (\alpha ,\beta )=\frac{1}{\pi }\left\{\cos ^{-1}\left(\frac{\sin 
\alpha } {\sin \beta }\right)-\cos ^{-1}\left(\frac{\tan \alpha } 
{\tan \beta }\right)\cos \beta \right\},
\end{equation}
where $\alpha =\Phi $/2 and $\beta $ is the half-angle defining the 
proportion of the surface for which each assistant makes a definite reading 
(zero corresponds to none; $\pi $/2 to the whole surface). 
$P_{SS}$ achieves 
a maximum of $\frac{1}{2}(1 - \cos \beta)$
when 
$\alpha $ = 0, which is less than the QM prediction of 0.5 unless $\beta $ 
is $\pi $/2. When $\alpha \ge \beta $, it is zero (see Fig.~\ref{fig7}).

The presence of the zero region has the interesting consequence that if 
$\beta $ is too small (less than $\pi $/4) the derived quantum correlation 
is undefined for a region in the neighbourhood of $\alpha = \pi/4$ or 
$\Phi =\pi/2$. So far as actual experiments go, however, the matter is 
largely academic, since there are always background ``dark counts'' 
and other ``accidentals'' that ensure that the observed counts are never 
zero. In the actual experiments the whole curve would be smoother, as hard 
divisions between regions scoring 1 and those scoring zero would be unlikely 
to occur.

\begin{figure}[htbp]
\centerline{\includegraphics[width=2.8in,height=2.0in]{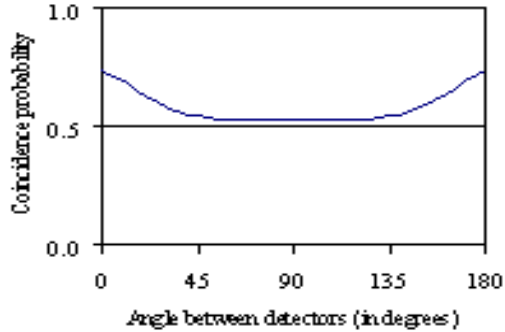}}
\caption{Total coincidence rate, $T_{obs}/N.$}
\label{fig8}
\end{figure}

To obtain the prediction for ``unlike'' coincidences such as {\it NS}, we replace 
$\alpha$ by ($\pi/2 - \alpha$). We can now find the predicted value (Fig.~\ref{fig8}) 
for the total observed coincidence rate, $T_{obs} /N = (NN + SS - NS - SN)/N$.

The fact that this is not constant was recognised by Pearle in his 1970 
paper, and can be used as a test for the validity of the QM model. It is, 
however, not a conclusive one, as can be seen by consideration of the 
situation in which one detector is perfect and the other has missing bands. 
The estimated quantum correlation could violate Bell's limit despite the 
fact that $T_{obs}$ was constant. An important fact (also noted by Marshall 
\etal~\cite{Marshall:1983}) is that the values predicted for the 
``Bell test angles'' of $\pi$/4 and 3$\pi$/4 are equal.

We can derive the estimate of the ordinary ``normalised'' quantum 
correlation in which division is by $T_{obs}$ , with results as shown in Fig. 
5, but it is of interest to look also at the ``unnormalised'' one, 
corresponding to division by $N$:
\begin{displaymath}
   P_{NN} + P_{SS} - P_{NS} - P_{SN},
\end{displaymath}
\noindent plotted in Fig. 9. 

\begin{figure}[htbp]
\centerline{\includegraphics[width=2.93in,height=2.48in]{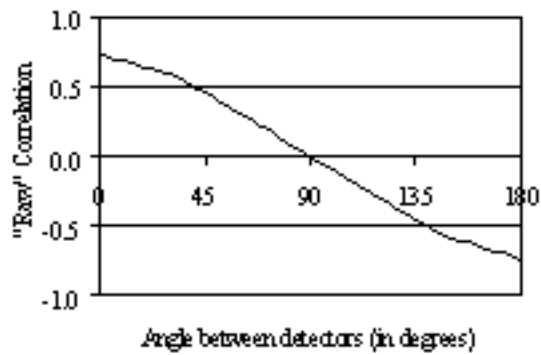}}
\caption{Unnormalised ``quantum correlation''.}
\label{fig9}
\end{figure}

The match with the QM prediction is considerably less impressive, the curve 
not reaching the maximum of 1 and not having the feature of a zero slope for 
parallel detectors ($\alpha $ = 0). Whilst (for the chosen example, with 
$\beta$ set at $75^{\circ}$) the model gives the CHSH69 test statistic of $S = 
3.331 > 2$, the unnormalised estimate will never exceed 2. The values at 
the ``Bell test angles'' will always all be numerically less than 0.5.

\section{Discussion}
We have confirmed by means of this counter-example Pearle's 1970 finding that 
the CHSH69 test as usually implemented (using the sum of observed coincidences 
in place of number of emitted pairs) rests on an assumption --- that of fair sampling 
--- that cannot be made lightly. Not only this, but the model demonstrates 
that local hidden variable theories (or what Clauser and Horne~\cite{Clauser:1974} 
prefer to call Objective Local Theories, since the hidden variables are not fully 
deterministic as originally defined) that violate Bell inequalities are not 
all ``weird''. They do not, as often stated (see for example Lalo\"{e}'s 
review article, ref.~\cite{Laloe:2001}) require ``conspiracies'' between the 
detectors. Furthermore, the model shows the fallacy of the belief, held by 
Bell and frequently quoted, that ``imperfections'' are unlikely to increase 
the significance of his tests (page 109 of his book, ``Speakable and 
Unspeakable''~\cite{Bell:1987}).

Bell himself, incidentally, was well aware that in order to get a strictly valid test 
it was necessary to know the number of pairs emitted by the source. He would have liked 
to see ``event-ready detectors'', counting emissions as they occurred. If 
this number were to be used as denominator in the estimates $E$, the test would 
be valid and the ``local realist'' model represented by the ball would {\it not} 
violate the inequality.  

In view the serious possibility of bias, the routine use of 
the CHSH test and assumption of fair sampling would seem to require explanation.
On reflection, several possible reasons spring to mind.

\begin{itemize}
\item Neither Pearle's nor later rediscoveries have 
succeeded in conveying at an intuitive level just {\it how }the detection 
loophole produces bias, or why it should not be taken for granted that experimenters, 
following proper experimental methods, would not have selected a fair sample.  
Without an understanding of how the loophole arises, it is not clear that the
sample chosen is not directly under the experimenter's control.
\item The use of event-ready detectors is not practical since we are dealing with 
``quantum'' events that are destroyed when observed (though see recent
proposals for ``loophole-free'' experiments that manage to circumvent this 
problem~\cite{Garcia:2004,Nha:2004}).
\item In almost all real experiments photons are used as the particles, 
and there exist to date no perfect detectors for ``single photons''.  The resulting
large numbers of non-detections mean that division by the number of emitted 
pairs would, even if a theory-free and agreed method of estimating this could be found,
produce ratios so small that there would be no possibility of infringement of any 
Bell test.  
\item To a quantum theorist who is convinced that light really does consist of photons 
(as opposed to merely being for some purposes modelled as such), since all photons
of a given frequency are necessarily identical it would seem impossible for them to 
possess any additional property (a component of its ``hidden variable'') that would 
allow the detector to discriminate between two that had different histories.  
Under a wave model of light, there is no such problem: each pulse has an intensity 
and this would be expected to be affected by passage through a polariser in such a way
as to form a link between the initial polarisation (the main component of the hidden 
variable), the detector setting and the probability of detection.
\end{itemize}

Is there any way of rescuing the situation?   As mentioned above, there is the possibility
of eventually conducting a loophole-free test, but in the meantime, though we can never prove
the sample is fair, our model (and, indeed, Pearle's) suggests a way of demonstrating
the fact when it is {\it not} fair.  Whenever, as assumed in our current model,
there is rotational invariance, a rigorous test should be conducted to check for 
variations in total coincidence counts ($T_{obs}$).  Here, as is clear from the present 
paper and also from Pearle's, it is important to check not only the ``Bell test angles''
but the full range, the greatest differences being expected at the intermediate angles.  

A completely different and arguably preferable alternative would be not to use the CHSH69
test but the CH74 one instead.  For this test Clauser and Horne used single-channel
polarisers (in the language of the chaotic ball model, they looked at just the 
$N$'s, say, ignoring the $S$'s).  The assumption that is part of the proof of the 
suspect CHSH69 or visibility tests --- that we have fair sampling, with the observed 
coincidences a representative sample of the emitted pairs --- does not enter into 
the question.

In practice there are important parameters --- in particular, beam intensity 
and characteristics of the photodetectors --- that are at the experimenter's 
discretion. Realist models, whether following the ball analogy or working 
directly from the basic theoretical formula (\ref{eq1}), suggest that the coincidence 
curve will be strongly influenced by the choices made. If we are genuinely 
trying to find the best model for the physical situation, is it not 
necessary to include all these relevant parameters? They play a natural part 
in the realist approach but in the quantum-mechanical one are ignored: in 
quantum mechanics, detectors are characterised by a single ``quantum 
efficiency''. 

It is generally assumed that classical and quantum theory agree on an 
important consequence of this characterisation: adherence of the probability 
of detection to Malus' Law, but, whether because the polarisers are 
imperfect or the detectors not exactly ``square law'', this adherence will 
not be exact (see, for example, refs.~\cite{Thompson:1997} and \cite{Thompson:1999a}). 
The actual relationship, and hence the observed coincidence curve, may not be 
quite sinusoidal. The majority of Bell test experiments have concentrated on 
just the few points needed to estimate the test statistic, not looking at 
enough data points to check the shape of the complete coincidence curve, let alone 
the constancy of the total coincidence count. Any deviation from the sine curve should be 
regarded as an indicator in favour of local realism.

\section{Other loopholes}
The detection loophole is, at least among professionals, well known, but the 
fact that it affects some versions of Bell's test and not others is perhaps 
less well understood. Different loopholes apply to different versions, for 
each version comes with its attendant assumptions. Some loopholes come very much under 
the heading of ``experimental detail'' and have, as such, little interest to 
the theoretician. If we wish to decide on the value to be placed on a Bell 
test, however, such details cannot be ignored.

\subsection{Subtraction of ``accidentals''}
Adjustment of the data by 
subtraction of ``accidentals'', though standard practice in many 
applications, can bias Bell tests in favour of quantum theory. After a 
period in which this fact has been ignored by some experimenters, it is now 
once again accepted~\cite{Tittel:1998}. The reader should be aware, though, 
that it invalidates many published results~\cite{Thompson:1999a}.

\subsection{Failure of rotational invariance}
The general form of a Bell test 
does not assume rotational invariance, but a number of experiments have been 
analysed using a simplified formula that depends upon it. It is possible 
that there has not always been adequate testing to justify this. Even where, 
as is usually the case, the actual test applied is general, if the hidden 
variables are not rotationally invariant, {\it i.e.}~if some values are favoured 
more than others, this can result in misleading descriptions of the results. 
Graphs may be presented, for example, of coincidence rate against $\Phi$, 
the difference between the settings $a$ and $b$, but if a more comprehensive set 
of experiments had been done it might have become clear that the rate 
depended on $a$ and $b$ separately~\cite{Thompson:1999b}. Cases in point may be 
Weihs {\it et al}'s 1998 experiment, presented as having closed the ``locality'' 
loophole~\cite{Weihs:1998}, and Kwiat {\it et al}'s demonstration of entanglement using 
an ``ultrabright photon source''~\cite{Kwiat:1999}.

\subsection{Synchronisation problems}
There is reason to think that in a few 
experiments bias could be caused when the coincidence window is shorter than 
some of the light pulses involved~\cite{Thompson:1997}. These include one of 
historical importance --- that of Freedman and Clauser, 
in 1972~\cite{Freedman:1972} --- which used a test not sullied by either of the 
above possibilities.

\subsection{``Enhancement''}
\label{no enhancement}
Tests such as that used by Freedman and Clauser 
(essentially the CH74 test) are subject to the assumption that there is ``no 
enhancement'', {\it i.e.}~that there is no hidden variable value for which the 
presence of a polariser {\it increases} the probability of detection. This 
assumption is considered suspect by some authors, notably Marshall and Santos, 
but in practice, in the few instances in which the CH74 inequality has been used, 
the test has been invalidated by other more evident loopholes such as the 
subtraction of accidentals.

\subsection{Asymmetry}
Whilst not necessarily invalidating Bell tests, the 
presence of asymmetry (for instance, the different frequencies of the light 
on the two sides of Aspect's experiments) increases the options for local 
realist models~\cite{Caser:1984}.

\subsection{Yet other loopholes}
A loophole that is notably absent from the above list is the so-called 
``locality'' or ``light-cone'' one, whereby some unspecified mechanism is 
taken as conveying additional information between the two detectors so as to 
increase their correlation above the classical limit. In the view of many 
realists, this has never been a serious contender. John Bell supported 
Aspect's investigation of it (see page 109 of ref.~\cite{Bell:1987}) 
and had some active involvement with the work, being on the examining board 
for Aspect's PhD. Weihs \etal improved upon the test in their experiment of 
1998~\cite{Weihs:1998}, but nobody has ever put forward plausible ideas for 
the mechanism. Its properties would have to be quite extraordinary, as it is 
required to explain ``entanglement'' in a great variety of geometrical setups, 
including over a distance of several kilometers in the Geneva experiments of 
1997-8~\cite{Tittel:1997, Tittel:1998}.

There may well be yet more loopholes. For instance, in many experiments the 
electronics is such that simultaneous `+' and `--' counts from both outputs 
of a polariser can never occur, only one or the other being recorded. Under 
QM, they will not occur anyway, but under a wave theory the suppression of 
these counts will cause even the basic realist prediction 
(expression (\ref{eq1}) above) to involve ``unfair sampling''. The effect is 
negligible, however, if the detection efficiencies are low, since the 
three- or four-fold coincidences concerned (two on one side, one or more on 
the other) then hardly ever happen.

\section{Explaining "anomalies"}
Alain Aspect's set of three experiments is deservedly given pride of place 
in any listing of Bell test trials. His PhD thesis~\cite{Aspect:1983}, which 
gives very much more detail of the experiments than could be included in 
Physical Review Letters, shows evidence of careful reasoning and meticulous 
attention to detail. That detail includes, however, mention of some 
anomalies that perhaps deserve more attention. None reached the level of 
statistical significance, but Aspect evidently recognised that they were 
potentially important. 

One anomaly in particular the reader will recognise as giving cause for 
concern: the total number of coincidences in the experiment in which the 
CHSH69 test was used (the first 1982 one, using two-channel polarisers) 
varied slightly with detector setting, so the sample may not have been 
strictly ``fair''. Instead of increasing replication to check if the 
variations were consistent, though, Aspect derived (pp 124-7 of his thesis) 
a generalised Bell test designed to compensate for them. When applied to the 
1982 experiment, this modified test seemed to show that the effect 
associated with the anomalous total could not have caused significant bias. 
The new test is mentioned briefly in the relevant paper, but the promised 
publication on the subject does not appear to have materialised. Its 
validity can be challenged and, especially since Aspect's methods have been 
used as a precedent by others, the matter deserves wider publicity. 

Another minor anomaly in the two-channel experiment was a discrepancy 
between the counts of `+ --' and `-- +' coincidences. A slight adaptation of 
the Chaotic Ball model to allow for two known facts of the experiment --- 
asymmetry of $A$ and $B $ detectors related to the differences in wavelength, 
and slight bias in the polarisers --- shows straightforwardly how this might 
arise.

\section{Conclusion}
The ``Chaotic Ball'' models a hypothetical Bell test experiment in a manner 
that encourages the use of intuition and realism. It illustrates the fact, 
well known to those working in the field, that if not all ``particles'' are 
detected there is risk of bias in the tests used, which are no longer able 
to discriminate between the ``nonseparable'' quantum-mechanical model and 
local realism. Since the Bell tests themselves fail to discriminate, we 
suggest that more comprehensive experiments, covering more data points and a 
wider range of conditions, are desirable. The full predictions of the model, 
not just the Bell test values, should be checked. There is a serious 
possibility that quantum entanglement may never in fact happen.

The ball model is a special case of a general class of local realist models, 
many of which are covered by equation (\ref{eq1}) above --- essentially the same 
formula that Bell assumed, reflecting the local realist assumption that 
{\it the observed correlations originate from shared properties acquired at 
the source, and the detection events are independent}. 
The formula presupposes the standard case, in which the experiment consists 
of a series of readily identifiable events, a single particle pair being 
associated with each. If this is not the case, for example where there are 
``accidentals'' or where there is uncertainty in the time of detection, 
further generalisation is needed.

Note that the phrase in italics above is all we need in order to specify a 
local realist description. This simplicity is in stark contrast to the 
oft-stated objection (raised, e.g. by Lalo\"{e}, ref.~\cite{Laloe:2001}) that 
local realist models are complicated, ``weird'' and ``ad hoc''. Another frequent 
objection is that local realism cannot match quantum theory when it comes to 
accurate quantatitive predictions. True, it cannot easily match {\it exactly }
the quantum-mechanical coincidence formulae (the ball model, illustrating 
principles only, does not even attempt to do so), but what is required is surely a 
match with experimental results, not with the quantum theory predictions. We 
suggest that the quantum-mechanical predictions are not necessarily correct, 
and, indeed, that the apparently accurate matches observed are slightly 
illusary, being improved by several factors: (a) adaptations are made (as 
with the local realist model) to the formulae to allow for experimental 
conditions (b) various experimental parameters such as beam intensity and 
detector efficiency are at the experimenter's discretion, and 
(c) the full coincidence curve is often not investigated, the experiment 
covering only the minimum points needed for the Bell test.

Our underlying assumption when we look at the optical case (where we use not 
the deterministic ball model but the general ``stochastic'' model of 
equation (\ref{eq1})) is that the energy of each individual light pulse (``photon'') 
is split at the polariser --- something no photon can do. The intensity of 
the emerging pulse then influences statistically the probability of the 
detector firing. Despite the success of the photon model of light in many 
applications of quantum optics, this success is at the price of recognised 
conceptual difficulties. The possibility that wave models that allow for the 
idiosyncrasies of the apparatus used --- deviations from Malus' Law for 
example --- may be able to account with no such difficulties for {\it all} optical 
phenomena is the subject of ongoing research by one of us (CHT).

\ack
Thanks are due to Franck Lalo\"{e} for encouragement to air again the 
Chaotic Ball model. The work would not have been completed without the moral 
support of David Falla of the Physics Department of University of Wales, 
Aberystwyth, and of the many who have expressed appreciation of C H 
Thompson's web site or contributions to Internet discussions. Her use of the 
computer and library facilities at Aberystwyth was by kind courtesy of the 
Department of Computer Science, of which she was until December 2003 an 
associate member.

\end{document}